# DNN-based uncertainty estimation for weighted DNN-HMM ASR

*José Novoa, Josué Fredes, Nestor Becerra Yoma*
Speech Processing and Transmission Lab., Universidad de Chile
nbecerra@ing.uchile.cl

## Abstract

In this paper, the uncertainty is defined as the mean square error between a given enhanced noisy observation vector and the corresponding clean one. Then, a DNN is trained by using enhanced noisy observation vectors as input and the uncertainty as output with a training database. In testing, the DNN receives an enhanced noisy observation vector and delivers the estimated uncertainty. This uncertainty in employed in combination with a weighted DNN-HMM based speech recognition system and compared with an existing estimation of the noise cancelling uncertainty variance based on an additive noise model. Experiments were carried out with Aurora-4 task. Results with clean, multi-noise and multi-condition training are presented.

## 1. Introduction

Uncertainty variance in noise removal was proposed initially to weight the information provided by frames according to their reliability in DTW and HMM algorithms [1] [2] [3]. To achieve this purpose, the enhanced features (e.g. MFCC or filter-bank log-energies) should be considered random variables with the corresponding mean and variance. According to [4], it was proposed "the replacement of the ordinary output probability with its expected value if the addition of noise is modelled as a stochastic process, which in turn is merged with the hidden Markov model (HMM) in the Viterbi algorithm." Consequently, the new output probability for the generic case of a mixture of Gaussians can be seen as the definition of a stochastic version of the weighted Viterbi algorithm. This is due to the fact that the final variances of the Gaussians correspond to the sum of the HMM and uncertainty variances. If the uncertainty variances rise, the discriminability of the GMM observation probability is reduced and the decoding process depends more on the language model [5]. The Viterbi decoding algorithm, which incorporates the uncertainty in noise cancelling is denoted as Stochastic Weighted Viterbi (SWV) algorithm because the growth of the GMM variances leads to a discriminability decrease of those frames with high uncertainty. Results with GMM-HMM-based speaker verification [4] and speech recognition [5] [6] suggested that SWV can provide significant WER reductions when speech signals are corrupted with additive, convolutional and coding-decoding distortion.

A similar result was later obtained in [7] by marginalizing the joint conditional pdf of the original and corrupted cepstral features over all possible unseen clean speech cepstra. As an alternative of using a model for additive noise, as in [4], the pdf of the noisy features, given the clean coefficients, was assumed to be as a Gaussian distribution. Nevertheless this result employed the same idea of uncertainty proposed in [1] [2] [3]. Moreover, in [7], the weighting nature of the use of uncertainty was not analysed. In [8], a new classification rule was described by proposing an integration over the feature space instead of over the model-parameter space. It was evaluated with connected speech recognition. The enhancement uncertainty variances were estimated by using a probabilistic and parametric model of speech distortion. Two adaptation schemes were proposed to preserve the observation uncertainty in [9]. The results were achieved with connected digits. It is worth emphasizing that in [5] and [6] a generalization of the model presented in [4] was successfully applied to a continuous speech recognition task.

The uncertainty estimation of speech features was also tackled later in [10] [11] [12] [13]. More specifically in [13], it was shown that short-term Fourier transform (STFT) uncertainty propagation can be combined with the Wiener filter to compute minimum mean square error (MMSE) estimations in the feature domain for several parameter extraction methods. On the other hand, despite the noise cancelling uncertainty being presented only for band-pass filters and MFCC coefficients, the proposed modelling employed in [1] [2] [3] [4] does not require consideration of a Gaussian distribution for the additive noise in the STFT domain. Additionally, the non-linear log function is included by definition in the uncertainty estimation with spectral subtraction.

In the context of band-pass filter bank analysis based features, as mentioned above, the uncertainty in noise cancelling was proposed firstly in [1] [2], and further developed in [4]. According to [4] the uncertainty variance in noise cancelling in a band-pass filter is expressed as:

$$\text{Var}\left[\log\left(\overline{s_m^2}\big|\overline{y_m^2}\right)\right] = \begin{cases} \dfrac{2 \cdot c_m \cdot \text{E}[\overline{n_m^2}]}{\overline{y_m^2} - \text{E}[\overline{n_m^2}]} & , \text{if } \dfrac{\overline{y_m^2} - \text{E}[\overline{n_m^2}]}{10 \cdot c_m \cdot \text{E}[\overline{n_m^2}]} \geq 1 \\ -\dfrac{\overline{y_m^2} - \text{E}[\overline{n_m^2}]}{50 \cdot c_m \cdot \text{E}[\overline{n_m^2}]} + 0.4 & , \text{else} \end{cases} \quad (1)$$

where $\overline{s_m^2}$, $\overline{y_m^2}$ and $\text{E}[\overline{n_m^2}]$ are the estimated original clean energy, observed noisy energy and estimated noise energy at filter $m$, respectively. Also, $c_m$ is a correction coefficient that considers the short-term correlation between the clean and noise signals. According to [1], $\text{E}[\log(\overline{s_m^2}|\overline{y_m^2})] = \log(\overline{y_m^2} - \text{E}[\overline{n_m^2}])$, where $\overline{y_m^2} - \text{E}[\overline{n_m^2}]$ can be seen as the spectral subtraction (SS) estimate of the clean signal. According to [1] and [4], the uncertainty variance of the Mel filter bank and MFCC can be obtained with (1). The uncertainty variance of delta and delta-delta features can be estimated as in [4]. This uncertainty variance is a key component of the SWV algorithm, which can lead to significant improvements in HMM-based speaker verification and speech recognition tasks.

Uncertainty propagation has been addressed by several authors in the last few years. Various uncertainty-of-observation (UoO) methods have been developed by extending the idea of using the uncertainty in noise cancelling to modify the acoustic model probability [7] [10] [14] [15] [16] [17]. The main motivation is the same as the one in SWV by considering the enhanced features as random variables, rather than estimated coefficients. Thus, the uncertainty incorporated by the enhancement process is considered the variance of the obtained feature. These random variables are then analytically propagated and modify the variance of the acoustic model. Nevertheless when applying this strategy to a DNN-based system, the problem of uncertainty propagation cannot be analytically treated without important approximations. Due to the fact that a DNN is not a probabilistic model it is not clear how to modify the acoustic pseudo-likelihood given the feature uncertainty. Some techniques for uncertainty propagation, such as the unscented transform (UT) [18] and piecewise exponential approximation (PIE) [19], have been proposed. UT corresponds to a method for propagating the statistics of a random variable through a nonlinear transformation. A set of *sigma points* is deterministically selected to represent the distribution of the random variable. Next, these points are propagated employing a given nonlinear function, the DNN in this case, and the mean and variance of the transformed set are estimated. This method differs from the Monte Carlo approach in that no random samples are required, and only a low number of points is needed [20].

In a DNN-HMM ASR system, the DNN delivers a pseudo-log-likelihood defined as [21]:

$$\log[p(x_t|q_t = s)] = \log[p(q_t = s|x_t)] - \log[p(s)] \quad (2)$$

where $x_t$ is the acoustic observation at time $t$, which corresponds to a window of input feature frames. Also, $q_t$ denotes one of the states or senones, $s \in [1, S]$, $S$ is the number of states or senones, and $p(s)$ is the prior probability of state $s$. The final decoded word string, $\widehat{W}$, is provided by [21]:

$$\widehat{W} = \arg\max_{W} \{\log[p(X|W)] + \lambda \cdot \log p(W)]\} \quad (3)$$

where $X$ denotes the sequence of acoustic observations $x_t$, and $p(X|W)$ is the acoustic model probability that depends on the pseudo log-likelihood delivered by the DNN, $\log[p(x_t|q)]$. Furthermore, $p(W)$ is the language model probability of word string $W$ and $\lambda$ is the constant that is employed to balance the acoustic model and language model scores. In [22] it was proposed a modification of the DNN-HMM decoding process by incorporating an uncertainty weight, $UW$, in (3):

$$\widehat{W} = \arg\max_{W} \{UW \cdot \log[p(X|W)] + \lambda \cdot \log[p(W)]\} \quad (4)$$

where $UW$ is defined for each $x_t$, i.e. $UW[x_t]$. Therefore, $UW[x_t] \rightarrow 0$ if the uncertainty of frames in $x_t$, as used in DNN-HMM systems, is high. Moreover, $UW[x_t] \rightarrow 1$ if the uncertainty of $x_t$ is low. Given $x_t$, DNN estimates $S$ pseudo-log-likelihoods, $\log[p(x_t|q = s)]$. At each $x_t$, the dispersion of $\log[p(x_t|q = s)]$ is defined as its variance computed over all the possible states or senones $s$. The weighted pseudo-log-likelihoods are expressed as $\mathcal{L}_w = UW[x_t] \cdot \log[p(x_t|q)]$. As a result, $\mathcal{L}_w$ has a lower variance or dispersion than $\log[p(x_t|q = s)]$ when $UW[x_t] < 1$. Note that the closer $UW[x_t]$ is to zero, the less dispersed is the distribution of $\mathcal{L}_w$. As a consequence, the information delivered by the acoustic model loses discriminability and the decoding process tends to rely more on the language model than on the acoustic model. On the other hand, if the uncertainty associated to $x_t$ is low, i.e. $UW[x_t]$ tends to be one, (3) is reduced to (4).

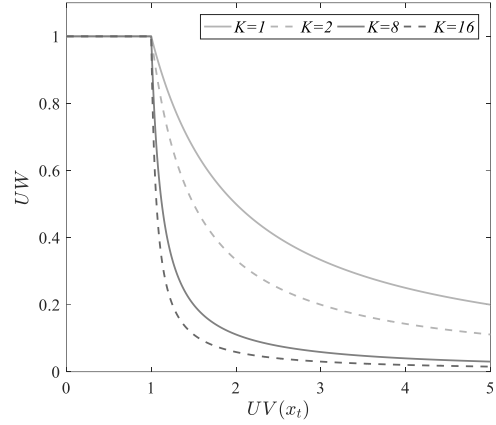

*Figure 1: Uncertainty weighting function [22]. Th was made equal to 1 and K was made equal to 1,2, 8 and 16.*

The motivation of the weighted DNN-HMM Viterbi algorithm as defined in (4) is to estimate and employ the uncertainty variance associated with the acoustic observation by providing an alternative technique to the uncertainty propagation methodology, which in turn requires many assumptions and approximations in the DNN framework. It should be noted that the use of the uncertainty variance at the DNN output had remained unsolved.

In this paper we make use of the weighting function proposed in [22], which is a generalization of the function presented in [1] [2] [3]:

$$UW[x_t] = \begin{cases} 1 & \text{, if } UV[x_t] \leq Th \\ \dfrac{Th}{K(UV[x_t] - Th) + Th} & \text{, if } UV[x_t] > Th \end{cases} \quad (5)$$

where $UV[x_t]$ is the uncertainty variance assigned to the acoustic observation, $x_t$; $Th$ is a threshold and $K$ is a constant that need to be tuned. Figure 1 depicts the weighting function defined in (5) with $Th = 1$ and several values of $K$.

## 2. MSE uncertainty

In this paper, instead of using any model as in (1), we estimate the uncertainty as the mean square error between a given enhanced noisy observation vector, $O_t^E$, and the corresponding clean one, $O_t^C$:

$$UV_t = \frac{1}{NS}\sum_{n=1}^{NS}(O_{t,n}^C - O_{t,n}^E)^2 \quad (6)$$

where t is the time index and NS is the number of static features, i.e. Mel filter bank (MelFB). If $2L + 1$ is the size of the window of input frame $x_t$ in the DNN, $UV[x_t]$ can be made equal to the averaged uncertainty variance within $x_t = [O_{t-L}^E, \cdots O_t^E, \cdots, O_{t+L}^E]$:

$$UV[x_t] = \frac{1}{(2L + 1)} \cdot \sum_{t-L \leq l \leq t+L} UV_l \quad (7)$$

Because the delta and delta-delta features are linear combinations of the static parameters, they were not included in (7).

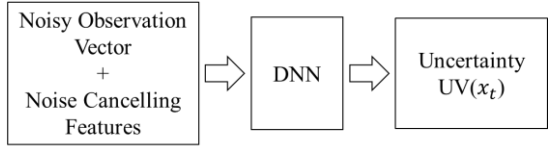

*Figure 2: Proposed DNN-based uncertainty estimation as defined in (6).*

In this paper we propose the estimation of $UV_t$ with neural networks (Fig. 2). First, a DNN is trained with enhanced noisy observation vectors as input and the uncertainty defined in (6) as output with a training database. Alternatively, the noisy observation vectors and other features such as noise cancelling parameters (e.g. the uncertainty variance as defined in (1)) can also be employed as input of the neural network. In testing, the DNN receives an enhanced noisy observation vector, or any other feature, and delivers the estimated uncertainty. This uncertainty is employed in combination with the weighted DNN-HMM based speech recognition system proposed in [22] according to (4) and (5).

## 3. Baseline system

The ASR experiments were performed on the Aurora-4 corpus [23] by using the Kaldi Speech Recognition Toolkit [24]. Three training sets from Aurora-4 were employed: the clean, multi-noise, and multi-conditions. Each training set contains 7,137 utterances from 83 speakers. The clean training set contains only clean data recorded with a Sennheiser HMD-414 microphone. The multi-noise set contains clean (25%) and artificially-degraded utterances (75%) with one out of six noises added at SNRs between 10 and 20 dB. They were recorded with the Sennheiser HMD-414 microphone. Half of the multi-condition training set was recorded with the Sennheiser HMD-414 microphone, while each utterance of the other half was recorded with one out of 18 different microphones, with noise added as in the multi-noise data. The testing and development databases were composed of 14 test sets clustered in four groups according to [23]: A, the original clean set; B, six sets corrupted with different additive noises; C, the clean set recorded with different microphones, and D, six sets corrupted with different additive noises and microphones. Each noisy test set contains 330 or 166 of artificially degraded utterances with one out of six noises added at SNRs between 5 and 15 dB. The development database was composed of 14 sets with 330 utterances each, clustered in four groups. The speakers and transcriptions in the development database are different from the testing ones. The development database was employed to avoid overfitting in the DNN training.

Spectral subtraction (SS) [25] was applied on a frame-by-frame basis to multi-noise and multi-condition training sets, and to testing data. The compensated Mel filter $m$ is defined as:

$$FE_m^{SS} = \max\{\beta \times FE_m \,;\, FE_m - \alpha(SNR_m) \times E[\overline{n_m^2}]\} \quad (8)$$

where

$$\alpha(SNR_m) = \begin{cases} \alpha_0 & , if\ SNR_m = 0dB \\ \alpha_0 - (\alpha_0 - 1) \cdot \frac{SNR_m}{18} & , if\ 0 < SNR_m < 18dB \\ 1 & , if\ SNR_m \geq 18dB \end{cases} \quad (9)$$

where $E[\overline{n_m^2}]$ is the noise energy estimated in non-speech intervals, as defined above; $FE_m$ is the filter energy without SS; $FE_m^{SS}$ is the compensated filter energy obtained with SS; $SNR$ corresponds to the segmental signal-to-noise ratio, where a segment corresponds to a frame; and $\beta$ defines a positive lower bound to the compensated filter energy. In this paper, $\alpha_0$ and $\beta$ are equal to 2.0 and 0.1, respectively. The uncertainty variances for the log energies of the Mel filters, and those for the corresponding delta and delta-delta features, were estimated as in [4] and as described in the Section I. Constant $c_m$ in (1) was made equal to 0.15 in all filters, as in [5]. Note that $c_m$ is merged in the weighting function, which in turn is defined by $K$ and $Th$. The feature vector was composed of 40 Mel filter bank (MelFB) features, and delta and delta-delta dynamic features, with consideration of an 11-frame context window. In a previous optimization step, the DNN-HMM baseline system with multi-condition training was tested with 24, 32, 40 and 56 MelFB filters. The lowest WER, 10.9%, was found with 40 filters and 330 utterances per testing set. This baseline WER is competitive with those published in the literature for the same task [26] [27] [28] [29].

## 4. Oracle results

To assess the potential of this approach we carried out experiments with group B of Aurora-4 database. Group B is composed of six subsets of noisy data that was generated by adding six types of noise to a set of clean utterances. The idea was to estimate the highest possible improvement in WER we could achieve with the weighted decoding and the uncertainty defined as in (4). First, the MSE uncertainty according to (6) was directly computed with each utterance in group B, noisy utterances, and the corresponding clean utterance. Then, parameters $Th$ and $K$ of the weighting function in (5) were tuned. The WERs obtained in this tuning are shown in Fig. 3. According to Fig. 3, with clean signal a dramatic reduction in WER equal to 30.2% (K=5, Th=8) was achieved when compared with the baseline with SS. As expected, with multi-noise and multi-condition training the reduction was much lower and equal to 2.15% (K=1, Th=12) and 1.82% (K=5, Th=18), respectively. The WER obtained with clean training is around twice as higher than the baseline with multi-noise or multi-condition training (see Table 2). If the uncertainty could be estimated accurately, a very descent accuracy could be achieved without any knowledge about the additive noise. This approach could also be interesting when the testing environment contains a distortion that was not seen during the multi-noise or multi-style training.

## 5. Experiments and discussion

The DNN based estimation of the MSE uncertainty, according to Section 2, was initially evaluated by implementing four network topologies. Each DNN was built using five hidden layers. The following configurations regarding the number of units per layer were evaluated: C1, 40-40-20-40-40; C2, 80-80-40-80-80; C3, 40-40-40-40-40; and, C4, 80-80-80-80-80. Moreover, three feature inputs (Fig. 2) were defined by employing the log of the normalized energy with respect the highest frame energy within the utterance and the uncertainty estimated according to (1) averaged over the 40 static MelFB features: f1, the log-normalized-frame-energy concatenated with the noisy features; f2, the log-normalized-frame-energy and average uncertainty concatenated with the enhanced features with SS as in (8) and (9); and, f3, the log-normalized-frame-energy concatenated with the enhanced features with SS as in (8) and (9). The DNNs were trained by making use of the Matlab Neural Network Toolbox with the Aurora-4 multi-noise training set using the MSE criterion and considering 20 iterations. Internally, the toolbox divides the data in training (70%), validation (15%) and testing (15%). The reference was the MSE uncertainty defined in (6) between the noisy and corresponding clean utterances. As can be seen in Table 1, the lowest MSE was obtained with DNN configuration C1 and input f2.

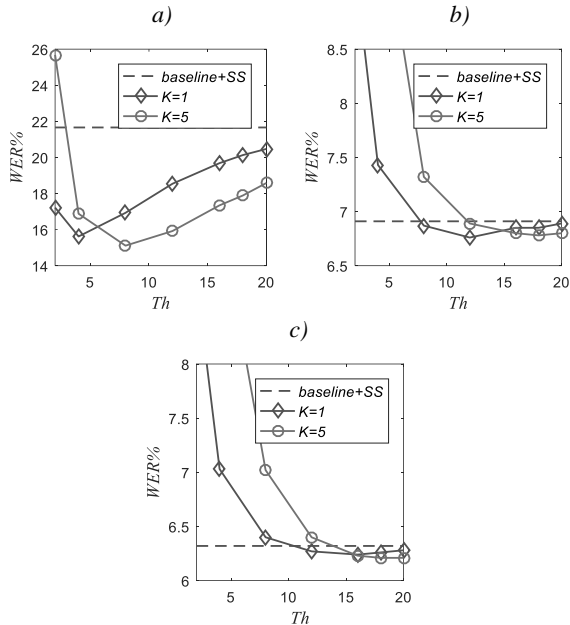

*Figure 3: Oracle WER v/s Th and K in (5) obtained with the test_166 sets from the Aurora-4 B group: a) clean training; b) multi-noise training; and, c) multi-condition training.*

The ASR experiments were carried out with the 14 166-utterance sets from Aurora-4. Results are shown in Table 2: Baseline+SS is the base line with SS; UW+UV SS_Model, corresponds to the weighted Viterbi algorithm as in (4) where the uncertainty variance is estimated as in [22] with (1); and, UW+UV DNN denotes the weighted Viterbi algorithm as in (4) with the DNN-based estimation of the MSE uncertainty proposed here. As can be seen in Table 2, with clean training the proposed method, UW+UV DNN, achieved an average reduction in WER equal to 9.8% (k=5, th=4) when compared with the baseline with SS, *baseline-SS*. In particular with group B, additive noise, UW+UV DNN led to a WER that is not as low as the oracle result in Section 4. Nevertheless, it is still 5.6% lower than the one provided with the weighted Viterbi algorithm where the uncertainty variance is estimated with the additive noise model according to (1). As expected, no improvements were reported with multi-noise and multi-condition training, but also no degradation was incorporated. However, as mentioned above, this approach should be applicable to situations when the testing environment contains distortion that was not seen during the multi-noise or multi-style training. Moreover, the results reported here suggest that the proposed uncertainty estimation and weighting scheme can outperform the existing propagation strategy.

*Table 1: MSE obtained in the DNN training process.*

|  |  | DNN Topology | | | |
| --- | --- | --- | --- | --- | --- |
|  |  | C1 | C2 | C3 | C4 |
| Feature set | f1 | 11.99 | 11.39 | 12.08 | 22.90 |
|  | f2 | **9.19** | 9.69 | 9.32 | 9.81 |
|  | f3 | 10.99 | 10.52 | 10.25 | 12.07 |

*Table 2: Summary of results from Aurora-4 test sets.*

| Training | Test | Baseline | Baseline +SS | UW+UV SS_model | UW+UV DNN |
| --- | --- | --- | --- | --- | --- |
| Clean | A | 2.21 | 2.50 | 2.65 | 2.47 |
|  | B | 29.34 | 21.72 | 19.80 | 18.69 |
|  | C | 20.92 | 21.62 | 20.07 | 19.71 |
|  | D | 49.76 | 42.70 | 40.58 | 39.38 |
|  | AVG | 35.55 | 29.33 | 27.50 | 26.47 |
| Multi Noise | A | 2.95 | 2.47 | 2.50 | 2.47 |
|  | B | 7.37 | 6.91 | 7.05 | 6.91 |
|  | C | 16.02 | 14.62 | 14.66 | 14.62 |
|  | D | 26.46 | 25.59 | 25.53 | 25.59 |
|  | AVG | 15.85 | 15.15 | 15.19 | 15.15 |
| Multi Condition | A | 3.43 | 3.13 | 3.09 | 3.13 |
|  | B | 6.26 | 6.32 | 6.42 | 6.32 |
|  | C | 8.07 | 7.07 | 7.03 | 7.07 |
|  | D | 17.63 | 17.75 | 17.52 | 17.75 |
|  | AVG | 11.06 | 11.04 | 10.98 | 11.04 |

## 6. Conclusions

In this paper we proposed a definition for the uncertainty in noise cancelling and a DNN based method to estimate it. No model for uncertainty noise cancelling or propagation is required, and the scheme proposed here can be employed in combination with any front end or distortion removal scheme. This new uncertainty is applied in combination with a DNN-HMM based weighted Viterbi algorithm. Special attention was focused on optimizing our DNN-HMM baseline system to produce a baseline WER that is competitive with those published elsewhere.

With clean training a dramatic oracle accuracy improvement was achieved. Nevertheless, with multi-noise and multi-condition training a low oracle reduction in WER was observed. In these matched training-testing conditions the accuracy of the DNN response would not depend significantly on the uncertainty in noise cancelling, and the weighted decoding would lose its effectiveness. However, if the accuracy of the DNN response is modelled with multi-noise and multi-condition training, this information can still be employed in combination with the scheme proposed herein.

It is worth emphasizing that the proposed method was tested with spectral subtraction, which only removes additive noise. Consequently, the combination of the DNN-based uncertainty estimation with distortion removal techniques that also account for channel mismatch and the modelling of the DNN response accuracy with multi-noise and multi-condition training, are proposed for future research.

## 7. Acknowledgements

The research reported here was funded by grants Conicyt-Fondecyt 1151306 and ONRG N62909-17-1-2002.